\documentclass[aps,prl,reprint,showpacs,superscriptaddress,longbibliography,floatfix]{revtex4-2}
\usepackage{mathtools,amssymb,graphicx,units}
\usepackage{amsmath,bm}
\usepackage{bbold}

\usepackage[plainpages=false,pdfpagelabels,colorlinks=true,linkcolor=red,urlcolor=blue,citecolor=blue,pdftitle={Title},pdfauthor={},pdfdisplaydoctitle=true,pdfduplex=DuplexFlipLongEdge]{hyperref}
\usepackage{verbatim}
\usepackage[english]{babel}
\usepackage{xcolor}
\usepackage{dsfont}

\usepackage[normalem]{ulem}

\usepackage{braket}
\usepackage{bbm}

\newcommand{\ketbra}[2]{\mathinner{|{#1}\rangle \langle{#2}|}}

\hypersetup{
           breaklinks=true,   % splits links across lines
           colorlinks=true,   % displays links as colored text instead of blocks
           pdfusetitle=true,  % \title and \author values into pdf metadata
        }

\begin{document}

\title{Diagnosing thermalization dynamics of non-Hermitian quantum systems via GKSL master equations}

\author{Yiting Mao}
\affiliation{Beijing Computational Science Research Center, Beijing 100084, China}

\author{Peigeng Zhong}
\affiliation{Beijing Computational Science Research Center, Beijing 100084, China}

\author{Haiqing Lin}
\email[]{haiqing0@csrc.ac.cn}
\affiliation{Beijing Computational Science Research Center, Beijing 100084, China}
\affiliation{School of Physics, Zhejiang University, Hangzhou, 310058, China}

\author{Xiaoqun Wang}
\email[]{xiaoqunwang@zju.edu.cn}
\affiliation{School of Physics, Zhejiang University, Hangzhou, 310058, China}

\author{Shijie Hu}
\email[]{shijiehu@csrc.ac.cn}
\affiliation{Beijing Computational Science Research Center, Beijing 100084, China}
\affiliation{Department of Physics, Beijing Normal University, Beijing, 100875, China}

\begin{abstract}
The application of the eigenstate thermalization hypothesis to non-Hermitian quantum systems has become one of the most important topics in dissipative quantum chaos, recently giving rise to intense debates. The process of thermalization is intricate, involving many time-evolution trajectories in the reduced Hilbert space of the system. By considering two different expansion forms of the density matrices adopted in the biorthogonal and right-state time evolutions, we have derived two versions of the Gorini-Kossakowski-Sudarshan-Lindblad master equations describing the non-Hermitian systems coupled to a bosonic heat bath in thermal equilibrium. By solving the equations, we have identified a sufficient condition for thermalization under both time evolutions, resulting in Boltzmann biorthogonal and right-eigenstate statistics, respectively. This finding implies that the recently proposed biorthogonal random matrix theory needs an appropriate revision. Moreover, we have exemplified the precise dynamics of thermalization and thermodynamic properties with test models.
\end{abstract}

\maketitle

Chaotic systems are known for their high sensitivity to initial conditions and their unpredictable long-term dynamics~\cite{Breuer_2007}. The famous Berry-Tabor~\cite{Berry_1977} and Bohigas-Giannoni-Schmit~\cite{Bohigas_1984} conjectures assert that classically integrable systems follow Poisson statistics of uncorrelated random variables, while chaotic quantum systems in the semiclassical limit exhibit statistics based on random matrix theory (RMT), which depends solely on symmetries rather than details of the models. For non-Hermitian chaotic systems, such as dissipative quantum systems described by Lindbladian, or non-unitary quantum dynamics governed by non-Hermitian Hamiltonians, or others~\cite{Sa_2020}, energy levels are predicted to follow Ginibre level statistics~\cite{Grobe_1988, Haake_1991, Akemann_2019} and to show notable cubic level repulsion~\cite{Grobe_1989}. Despite these established theories, the hidden relation between the dissipative quantum chaos~\cite{Altland1997, Kawabata_2019, Hamazaki2020} and thermalization in open quantum systems have been widely studied in recent years~\cite{JunjunXu27201, Li2021, Chan2022, Shivam2023, Da2023, li2024spectral}. 

\begin{figure}[t]
\begin{center}
\includegraphics[width=\columnwidth]{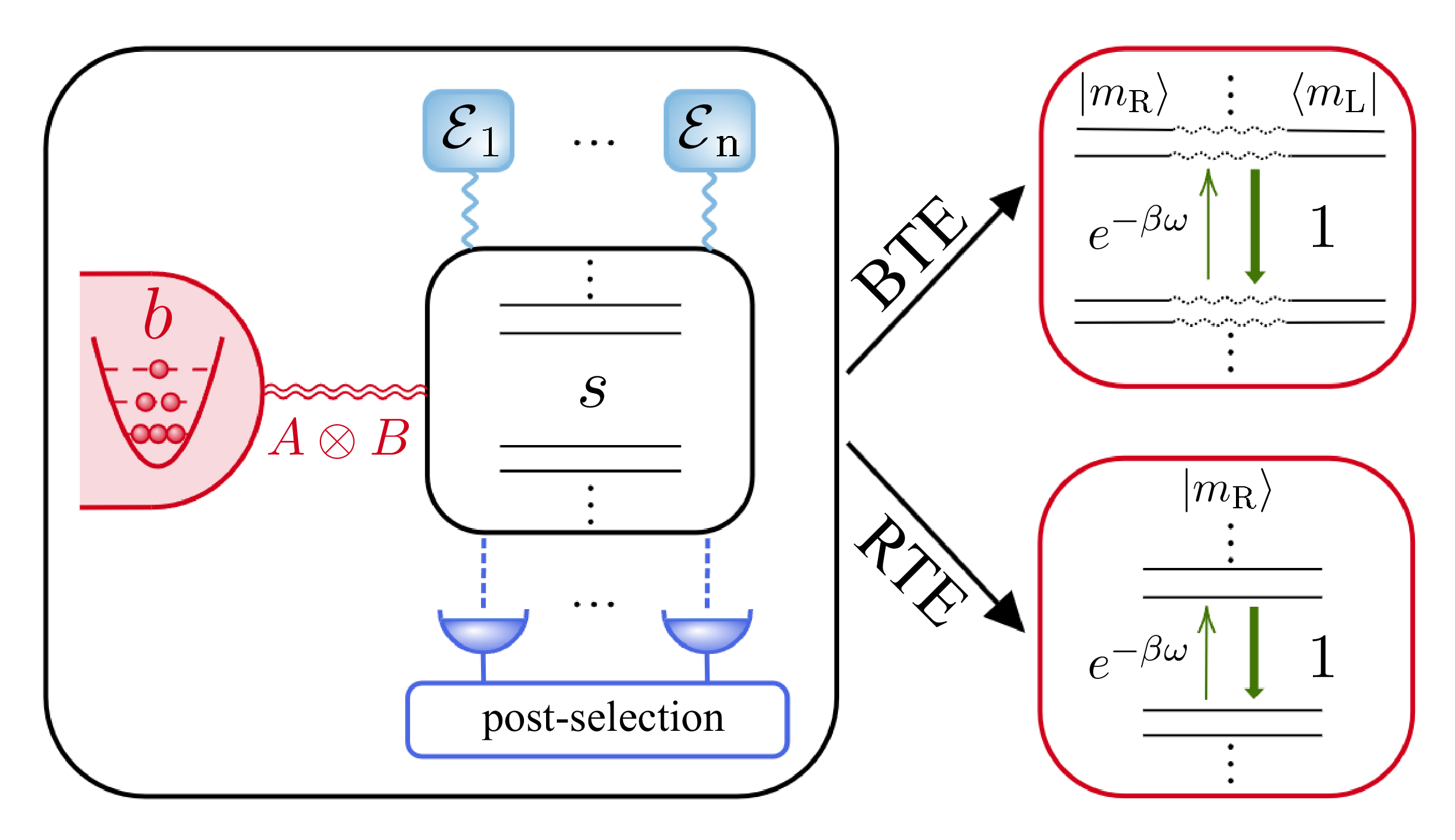}
\caption{\label{fig:fig1}
(Color online) When the system s is coupled to environments, it eventually reaches a stationary state described by either the Boltzmann biorthogonal statistics or the Boltzmann right-eigenstate statistics, depending on the time evolution assumption, BTE~\eqref{eq:rhoLR} or RTE~\eqref{eq:rhoRR}.
These environments, denoted as $\mathcal{E}$\textsubscript{1}, $\cdots$, $\mathcal{E}$\textsubscript{n}, introduce non-Hermitian effects due to the post-selection.
The bosonic heat bath remains in thermal equilibrium at temperature $T=1/\beta$ and interacts with the system through the coupling term $A \otimes B$.
Rightmost plots: In the statistics, the transitions between two levels with an energy difference of $\omega$ hold the detailed balance.
}
\end{center}
\end{figure}

The construction of the eigenstate thermalization hypothesis (ETH)~\cite{Deutsch_1991, Srednicki_1994, Srednicki_1999} to describe thermalization in general non-Hermitian systems has been a topic of debate in the past year originating from the non-Hermitian RMT and dissipative quantum chaos~\cite{Roy_2023, Cipolloni_2024}.
Metaphysically, the time evolution of the unit, consisting of a system and a heat bath, is governed by a non-Hermitian Hamiltonian $H$, with corresponding wave functions for the left eigenstate $\ket{\Phi(t)}$ and the right eigenstate $\ket{\Psi(t)}$.
Since the unit is not inherently ``closed", it is challenging to determine which bases are uniformly suitable for expanding the density matrix $\rho (t)$ of a state toward the long-term limit $t=+\infty$.
Among different choices evolving from the same initial density matrix $\rho(t=0)$: the biorthogonal time evolution (BTE) gives a mixed state with
\begin{eqnarray}\label{eq:rhoLR}
\rho(t) = \ketbra{\Psi(t)}{\Phi(t)} = e^{-i H t} \rho (0) e^{i H t}\, ,
\end{eqnarray}
and the right-state time evolution (RTE) yields a pure state with
\begin{eqnarray}\label{eq:rhoRR}
\rho(t) = \ketbra{\Psi(t)}{\Psi(t)} = e^{-i H t} \rho (0) e^{i H^\dag t}\, ,
\end{eqnarray}
as shown in Fig.~\ref{fig:fig1}.
The reference~\cite{Roy_2023} extended ETH to non-Hermitian systems using RTE~\eqref{eq:rhoRR} and proposed modifications in the off-diagonal measurement due to non-zero overlaps between different right eigenstates~\cite{Roy_2023}. The resulting Boltzmann right-eigenstate statistics (BRS) is in agreement with the previous works~\cite{Du_2022, Cao_2023_1, Cao_2023_2} on the Gorini-Kossakowski-Sudarshan-Lindblad (GKSL) master equation~\cite{Albash_2012, Mai_2013}.
In contrast, another study~\cite{Cipolloni_2024} argued that RMT violates ETH when evaluating the biorthogonal measurement of the operator under the assumption of BTE~\eqref{eq:rhoLR}. This argument is supported by significant fluctuations in both the diagonal and off-diagonal measurements as well as strong correlations between neighboring energy levels, which are absent in chaotic Hermitian systems~\cite{Cipolloni_2024}.
Although intriguing, to date, the biorthogonal chaotic dynamics in RMT have not yet been validated through model simulations.

In this work, we mainly study the thermalization process in the biorthogonal GKSL master equation using BTE~\eqref{eq:rhoLR}.
In contrast to RMT~\cite{Cipolloni_2024}, our results show that non-Hermitian systems consistently experience decoherence when the thermalization condition is satisfied, eventually leading to thermalization.
In the long term, the density matrix is shown to obey the Boltzmann biorthogonal statistics (BBS), which is further illustrated by two models.

\textit{Biorthogonal GKSL master equation.}
We start with a system (s) that is coupled to multiple environments (b, $\mathcal{E}$\textsubscript{1}, ..., $\mathcal{E}$\textsubscript{n}), one of which serves as a bosonic heat bath (b) in thermal equilibrium at temperature $T=1 / \beta$, as shown in Fig.~\ref{fig:fig1}.
The post-selection condition~\cite{Fukuhara_2013, MagaaLoaiza2014, Ashida_2017, Minganti2020} remains valid for other environments, such as those encountered in setups with superconducting qubits~\cite{Naghiloo_2019, Chen_2021} and optical lattices~\cite{Fukuhara_2013, Fukuhara_2015, Ashida_2017}.
The reduced Hilbert space of the unit (s plus b) $\mathcal{H}$ is a product of those for the system $\mathcal{H}_\text{s}$ (the dimension $d \ge 2$) and the heat bath $\mathcal{H}_\text{b}$, i.e., $\mathcal{H} = \mathcal{H}_\text{s} \otimes \mathcal{H}_\text{b}$.
The total Hamiltonian can be expressed as $H = H_\text{s} \otimes \mathbbm{1}_\text{b} + \mathbbm{1}_\text{s} \otimes H_\text{b} + H_\text{sb}$, where $H_\text{s}$, $H_\text{b}$ and $H_\text{sb}$ represent the Hamiltonians for the system, heat bath and their interaction, respectively.
In a general scenario without trivial symmetries, the non-Hermitian Hamiltonian $H^{\phantom{\dag}}_\text{s} \neq H_\text{s}^\dag$ has the left eigenstate $\ket{m_\text{L}}$ and the right eigenstate $\ket{m_\text{R}}$ for the $m$th non-degenerate state with energy $e_m$, satisfying both the biorthonormalization condition $\braket{m_\text{L} \vert n_\text{R}} = \delta_{mn}$ and the normalization condition $\braket{m_\text{R} \vert m_\text{R}} = 1$.
The coupling term can be simply written as $H_\text{sb} = A \otimes B$, where the operator $A$ defined in the system and the operator $B$ defined in the heat bath are Hermitian.
For the Hermitian system, the Hermiticity of the operators $A$ and $B$ is the key to preserving the energy conservation of the unit~\cite{Gaspard1999, deVega2017}.

Following BTE~\eqref{eq:rhoLR}, we get the Heisenberg equation in the interaction picture, i.e., $\text{d} \rho^\text{I} (t) / \text{d} t = -i [V (t),\, \rho^\text{I} (t)]$,
where the coupling term $V (t) = e^{i H_0 t} H_\text{sb} e^{-i H_0 t}$ and $\rho^\text{I} (t) = e^{i H_0 t} \rho (t) e^{-i H_0 t}$ are rotated by the Hamiltonian $H_0 = H_\text{s} \otimes \mathbbm{1}_\text{b} + \mathbbm{1}_\text{s} \otimes H_\text{b}$.
Assuming that the correlation time of the bath is much shorter than the relaxation time of the system when the Born-Markov approximation (BMA) is valid~\cite{Breuer_2007}, we get $\rho^\text{I} (t) \approx \rho^\text{I}_\text{s}(t) \otimes \rho^\text{I}_\text{b}(t)$ and $[V (t),\, \rho^\text{I} (0)] \approx 0$, where the density matrices in the interaction picture $\rho^\text{I}_\text{s} (t) = e^{i H_\text{s} t} \rho^{\phantom{\text{I}}}_\text{s} (t) e^{-i H_\text{s} t}$ and $\rho^\text{I}_\text{b}(t) = e^{i H_\text{b} t} \rho^{\phantom{\text{I}}}_\text{b} (t) e^{-i H_\text{b} t}$ are linked to the ordinary ones $\rho^{\phantom{\text{I}}}_\text{s} (t)$ and $\rho^{\phantom{\text{I}}}_\text{b} (t)$, respectively.
In addition, the heat bath always maintains its thermal equilibrium, i.e., $\rho^\text{I}_\text{b}(t) \approx \rho^\text{I}_\text{b} (0) = \bar{\rho}^{\phantom{\dag}}_\text{b} = e^{-\beta H_\text{b}} / Z_\text{b}$ with $Z_\text{b} = \text{tr}_\text{b} e^{-\beta H_\text{b}}$.
Consequently, one can obtain an equation:
\begin{eqnarray}
\label{eq:NHBMeq}
\frac{\text{d} \rho_\text{s}^\text{I} (t)}{\text{d} t} = -\int_0^\infty \text{d} \tau \ \text{tr}^{\phantom{\dag}}_\text{b} [V (t),\, [V (\tau),\, \rho^\text{I}_\text{s} (t) \otimes \bar{\rho}^{\phantom{\dag}}_\text{b}]]\, .
\end{eqnarray}
Using the biorthogonal bases $\ket{m_\text{L}}$ and $\ket{m_\text{R}}$, we get the expansion of $V (t) = \sum_{\omega \ne 0} e^{-i \omega t} A_\omega \otimes B (t)$ as convention~\cite{Breuer_2007}, where $A_\omega = \sum_{\omega_{mn} = \omega} A_{mn}$ sums over all selected pairs of $m$ and $n$ satisfying the level spacing $\omega_{mn} \equiv e_n - e_m = \omega$, and the operator for a pair of $m$ and $n$ is given by $A_{mn} = \mathbbm{A}_{mn} \ketbra{m_\text{R}}{n_\text{L}}$ with the expansion coefficient $\mathbbm{A}_{mn} = \braket{m_\text{L} \vert A \vert n_\text{R}}$.
We assume that the system has no trivial symmetries so that $\mathbbm{A}_{mn} \ne 0$ for any $m$ and $n$. In addition, $B (t) = e^{i H_\text{b} t} B e^{-i H_\text{b} t}$. In this work, we consider only the case of the real number $\omega_{m n}$, specifically in the $\mathcal{PT}$-unbroken region of the models.

We continue to use the definition of the correlation function for the bosonic heat bath~\cite{Breuer_2007}, i.e., $\Gamma(\omega) = \lim_{t \to +\infty} \int^t_0 \text{d} \tau \, e^{i \omega \tau} \text{tr}_\text{b} [B (t) B (t-\tau) \bar{\rho}^{\phantom{\dag}}_\text{b}] = \gamma (\omega) / 2 + i \eta (\omega)$, due to the Kubo-Martin-Schwinger (KMS) condition $\gamma(-\omega) / \gamma(\omega) = e^{-\beta \omega}$.
Within the rotating-wave approximation (RWA), only the zero-frequency microscopic processes provide the effective contribution~\cite{Fleming2010}.
We also ignore the imaginary part $\eta$, which only gives rise to the Lamb-shift term, affecting the evolution process instead of the steady state~\cite{Breuer_2007}.
At last, we get the biorthogonal GKSL master equation:
\begin{eqnarray}
\label{eq:GKSL}
\frac{\text{d} \rho_\text{s}}{\text{d} t} = \mathcal{L}_\text{s} [\rho_\text{s}] + \mathcal{L}_\text{b} [\rho_\text{s}] = \mathcal{L}_\text{s} [\rho_\text{s}] - \frac{1}{2} \mathcal{L}_\text{d} [\rho_\text{s}] + \frac{1}{2} \mathcal{L}_\text{j} [\rho_\text{s}]\, ,\ 
\end{eqnarray}
where
\begin{eqnarray}
\label{eq:bioGKSL}
\mathcal{L}_\text{s} [\rho_\text{s}] &=& -i [H_\text{s},\, \rho_\text{s}]\, ,\ \mathcal{L}_\text{d} [\rho_\text{s}] = \sum_{\omega \ne 0} \gamma (\omega) \{A_{-\omega} A_\omega,\, \rho_\text{s} \}\, ,\nonumber\\
\mathcal{L}_\text{j} [\rho_\text{s}] &=& \sum_{\omega \ne 0} \gamma (\omega) (A_\omega \rho_\text{s} A_{-\omega} +  A_\omega \rho_\text{s} A_{-\omega} )
\end{eqnarray}
represent the Liouvillians for the system, dissipation, and quantum jumps, respectively. And $[o^{\phantom{\dag}}_1,\, o^{\phantom{\dag}}_2]$ ($\{o^{\phantom{\dag}}_1,\, o^{\phantom{\dag}}_2\}$) denotes the regular (anti)commutator.

\textit{Thermalization condition}.
As we can see that the transition from the projector $\ketbra{m_\text{R}}{n_\text{L}}$ to the other one $\ketbra{(m^\prime)_\text{R}}{(n^\prime)_\text{L}}$ in Eq.~\eqref{eq:GKSL} conserves the energy, that is $\omega_{m^\prime m} = \omega_{n^\prime n}$, 
so that the density matrix $\rho_\text{s} (t)$ can be written as a direct sum of partial density matrices $\rho^{(\Delta)}_\text{s} (t)$ for the sectors labeled by the bias $\Delta= e_n - e_m$.

To assume that $M$ pairs of $m$ and $n$ construct
\begin{eqnarray}
\label{eq:pdensitymatrix}
\rho^{(\Delta)}_\text{s} (t) = \sum^M_{p=1} c_p (t) \ketbra{(m_p)_\text{R}}{(n_p)_\text{L}}\, ,
\end{eqnarray}
we get a Pauli master equation $\text{d} {\bm c} (t) / \text{d} t = \mathbbm{L} {\bm c} (t)$~\cite{Pauli28},
where a vector of ${\bm c} (t) = (c^{\phantom{\dag}}_1(t),\, \cdots,\,  c^{\phantom{\dag}}_M(t))^\intercal$ and two sets of ${\bm v}^{\phantom{\dag}}_m=\{m^{\phantom{\dag}}_1,\, \cdots,\, m^{\phantom{\dag}}_M\}$ and ${\bm v}^{\phantom{\dag}}_{n}=\{n^{\phantom{\dag}}_1,\, \cdots,\, n^{\phantom{\dag}}_M\}$ are introduced.
The detailed derivation is provided in the Supplementary Material.
In the time-independent matrix $\mathbbm{L}$, the diagonal elements $\mathbbm{L}_{pp} = -[\sum_{m^\prime \ne m_p} \gamma (\omega_{m^\prime m_p}) \kappa_{m_p m^\prime} + \sum_{n^\prime \ne n_p} \gamma (\omega_{n^\prime n_p}) \kappa_{n_p n^\prime}] / 2$ are generated by the dissipation $\mathcal{L}_\text{d}$ in Eq.~\eqref{eq:bioGKSL}, where symmetric $\kappa_{mn} = \mathbbm{A}_{mn} \mathbbm{A}_{nm}$ represents the transition rate, and both $m^\prime$ and $n^\prime$ run over all $d$ levels of the system.
We divide $\mathbbm{L}_{pp}$ into the part $-\sum_{q \ne p} \gamma (\omega_{m_q m^{\phantom{\dag}}_p}) (\kappa_{m_p m_q} + \kappa_{n_p n_q}) / 2$ from the sector $\Delta$, and the rest from $m^\prime \notin{\bm v}^{\phantom{\dag}}_m$ and $n^\prime \notin{\bm v}^{\phantom{\dag}}_n$. In contrast, the off-diagonal matrix element $\mathbbm{L}_{qp} = \gamma (\omega_{m_q m_p}) \mathbbm{A}_{m_q m_p} \mathbbm{A}_{n_p n_q}$ for $q \ne p$ is only determined by the quantum jumps $\mathcal{L}_\text{j}$ within the sector $\Delta$ as specified in Eq.~\eqref{eq:bioGKSL}. For a steady state ${\bm c}(t) = e^{\lambda t} \bar{\bm c}$, we get an equation $\mathbbm{L} \bar{\bm c} = \lambda \bar{\bm c}$ for the right eigenstate $\bar{\bm c}$ of $\mathbbm{L}$ with $\lambda$ representing the eigenvalue, as well as the one $\bar{\bm c}^\intercal_\text{L} \mathbbm{L} = \lambda \bar{\bm c}^\intercal_\text{L}$ for the corresponding left eigenstate $\bar{\bm c}_\text{L}$.

We have two scenarios to consider: (\textbf{i}) For the partial density matrix $\rho^{(0)}_\text{s} (t)$ in diagonal~\cite{BulnesCuetara2016} with $n_p \equiv m_p$ in Eq.~\eqref{eq:pdensitymatrix}, the equality $\sum^M_{q=1} \mathbbm{L}_{q p} = 0$ holds for any $p$, ensuring the existence of at least a stable left eigenstate $\bar{{\bm c}}_\text{L} = (1,\, \cdots,\,  1)^\intercal$ with the zero eigenvalue $\lambda=0$.
(\textbf{ii}) For the off-diagonal term $\rho^{(\Delta \ne 0)}_\text{s} (t)$, the possible presence of nonzero $\mathbbm{L}_{qp}$ indicates the existence of the level spacing degeneracy ($M \ge 2$), facilitating coherent transitions from $\ketbra{(m_p)_\text{R}}{(n_p)_\text{L}}$ to $\ketbra{(m_q)_\text{R}}{(n_q)_\text{L}}$.
Remarkably, if
\begin{eqnarray}
\label{eq:thermalizationCondition}
\mathbbm{A}_{m n} = (\mathbbm{A}_{n m})^*
\end{eqnarray}
is satisfied for arbitrary $m \ne n$, it is easy to prove that the matrix $\mathbbm{L}$ is strictly diagonally dominant because $\vert \mathbbm{L}_{pp} \vert  - \sum_{q \ne p} \vert \mathbbm{L}_{qp} \vert > \sum_{q \ne p} \gamma (\omega_{m_q m_p}) (\vert \mathbbm{A}_{m_p m_q} \vert - \vert \mathbbm{A}_{n_p n_q} \vert)^2 / 2 \ge 0$~\cite{Timm2009}.
In this case, the real parts of eigenvalues for sectors $\Delta \ne 0$ are always negative because of $\mathbbm{L}_{pp} < 0$, resulting in decoherence accordingly.
Overall, the condition~\eqref{eq:thermalizationCondition} is sufficient but unnecessary for thermalization in the long term, where only $\rho^{(0)}_\text{s} (t)$ survives.
For the convenience of the following discussion, we get a simple form of $\rho^{(0)}_\text{s} (t) = \sum^d_{m=1} c_m (t) \ketbra{m_\text{R}}{m_\text{L}}$, because $M = d$ is given for $\Delta=0$. It is noteworthy that Eq.~\eqref{eq:thermalizationCondition} is consistent with the stability condition in Ref.~\cite{Cao_2023_2} when $H_\text{s}$ is quasi-Hermitian.

\textit{Boltzmann biorthogonal statistics.}
When the thermalization condition~\eqref{eq:thermalizationCondition} is valid, one only considers the diagonal partial density matrix in the long term, with $\mathcal{L}_\text{s}[\rho_\text{s}^{(0)}]=0$ and
\begin{eqnarray}
\label{eq:bioLbath}
\mathcal{L} [\rho^{(0)}_\text{s}] = \sum_{m,n} \left( \gamma (\omega_{nm}) W_{mn} + \gamma (\omega_{mn}) W_{nm} \right)\, ,
\end{eqnarray}
where the transition superoperator is given by $W_{mn} = \kappa_{mn} (\ketbra{n_\text{R}}{n_\text{L}} - \ketbra{m_\text{R}}{m_\text{L}}) c_m (t)$ for indicating the probability transition from the $m$th level to the $n$th level.
The KMS condition establishes the detailed balance between those levels: $\gamma(\omega_{nm})/\gamma(\omega_{mn})=e^{-\beta \omega_{mn}}$, as shown in the upper rightmost plot of Fig.~\ref{fig:fig1}.
It is easy to get the right eigenstate $\bar{\bm c}=(\chi^{\phantom{\dag}}_1,\, \cdots,\, \chi^{\phantom{\dag}}_M)^\intercal$ for the stable steady state, where the Boltzmann weight for the $m$th level $\chi^{\phantom{\dag}}_m=e^{-\beta e_m} / Z_\text{s}$ and the partition function $Z_\text{s}=\sum_m e^{-\beta e_m}$ are given.
As a result, the system evolves to be a thermal state described by the density matrix for \textit{Boltzmann biorthogonal statistics}, i.e.,
\begin{eqnarray}
\label{eq:bioStatistics}
\bar{\rho}_\text{s}^\text{lr}=\sum^d_{m=1} \chi_m \ketbra{m_\text{R}}{m_\text{L}}\, .
\end{eqnarray}
In contrast, in the $\mathcal{PT}$-broken region, the perturbative quantum jumps and the dissipative terms in Eq.~\eqref{eq:bioGKSL} can be simultaneously ignored compared to the dissipation caused by the non-Hermiticity of the system due to Born-Markov approximation.
If $H_\text{s}$ has no degeneracy in the imaginary parts of the energy, the system cannot be thermalized as it evolves into the eigenstate corresponding to the energy with the maximal imaginary part, which has the highest growth rate of the amplitude and becomes dominant in the long term~\cite{Du_2022}.

\textit{Boltzmann right-eigenstate statistics.}
Under RTE~\eqref{eq:rhoRR}, we get the solution to Eq.~\eqref{eq:GKSL} similar to Ref.~\cite{Du_2022, Cao_2023_1, Cao_2023_2}, with Liouvillian defined as:
\begin{eqnarray}
\label{eq:rightGKSL}
\mathcal{L}_\text{s} [\rho_\text{s}] &=& -i [H_\text{s},\, \rho_\text{s}]_\dag\, ,\ \mathcal{L}_\text{d} [\rho_\text{s}] = \sum_{\omega \ne 0} \gamma (\omega) \{A_{-\omega} A_\omega,\, \rho_\text{s}\}_\dag\, ,\nonumber\\
\mathcal{L}_\text{j} [\rho_\text{s}] &=& \sum_{\omega \ne 0} \gamma (\omega) ( A_{\omega} \rho_\text{s} (A_\omega)^\dag + A_\omega \rho_\text{s} (A_\omega)^\dag )\, .
\end{eqnarray}
where $[o^{\phantom{\dag}}_1,\, o^{\phantom{\dag}}_2]_\dag = o^{\phantom{\dag}}_1 o^{\phantom{\dag}}_2 - o^{\phantom{\dag}}_2 o^\dag_1$ and $\{o^{\phantom{\dag}}_1,\, o^{\phantom{\dag}}_2\}_\dag=o^{\phantom{\dag}}_1 o^{\phantom{\dag}}_2 + o^{\phantom{\dag}}_2 o^\dag_1$ denote the modified commutator and anticommutator, respectively.

Analogous to BTE~\eqref{eq:rhoLR}, starting from an initial density matrix that can be written as the diagonal form of non-orthogonal self-normalized right eigenstates $\ket{m_\text{R}}$,
the thermalization condition~\eqref{eq:thermalizationCondition} yields a long-term stable thermal state with the density matrix for \textit{Boltzmann right-eigenstate statistics} (see details in the Supplementary Material), i.e.,
\begin{eqnarray}
\label{eq:rightStatistics}
\bar{\rho}_\text{s}^\text{rr} = \sum^d_{m=1} \chi_m \ketbra{m_\text{R}}{m_\text{R}}\, .
\end{eqnarray}
As shown in the lower rightmost plot of Fig.~\ref{fig:fig1}, the detailed balances can also be activated by the Liouvillian~\eqref{eq:bioLbath} with the redefined transition superoperator $W_{mn} = \kappa_{mn} (\ketbra{n_\text{R}}{n_\text{R}} - \ketbra{m_\text{R}}{m_\text{R}}) c_m (t)$.
Nevertheless, the condition~\eqref{eq:thermalizationCondition} can not guarantee that the off-diagonal terms in the density matrix disappear in the long term in general.

\begin{figure}[t]
\begin{center}
\includegraphics[width=\columnwidth]{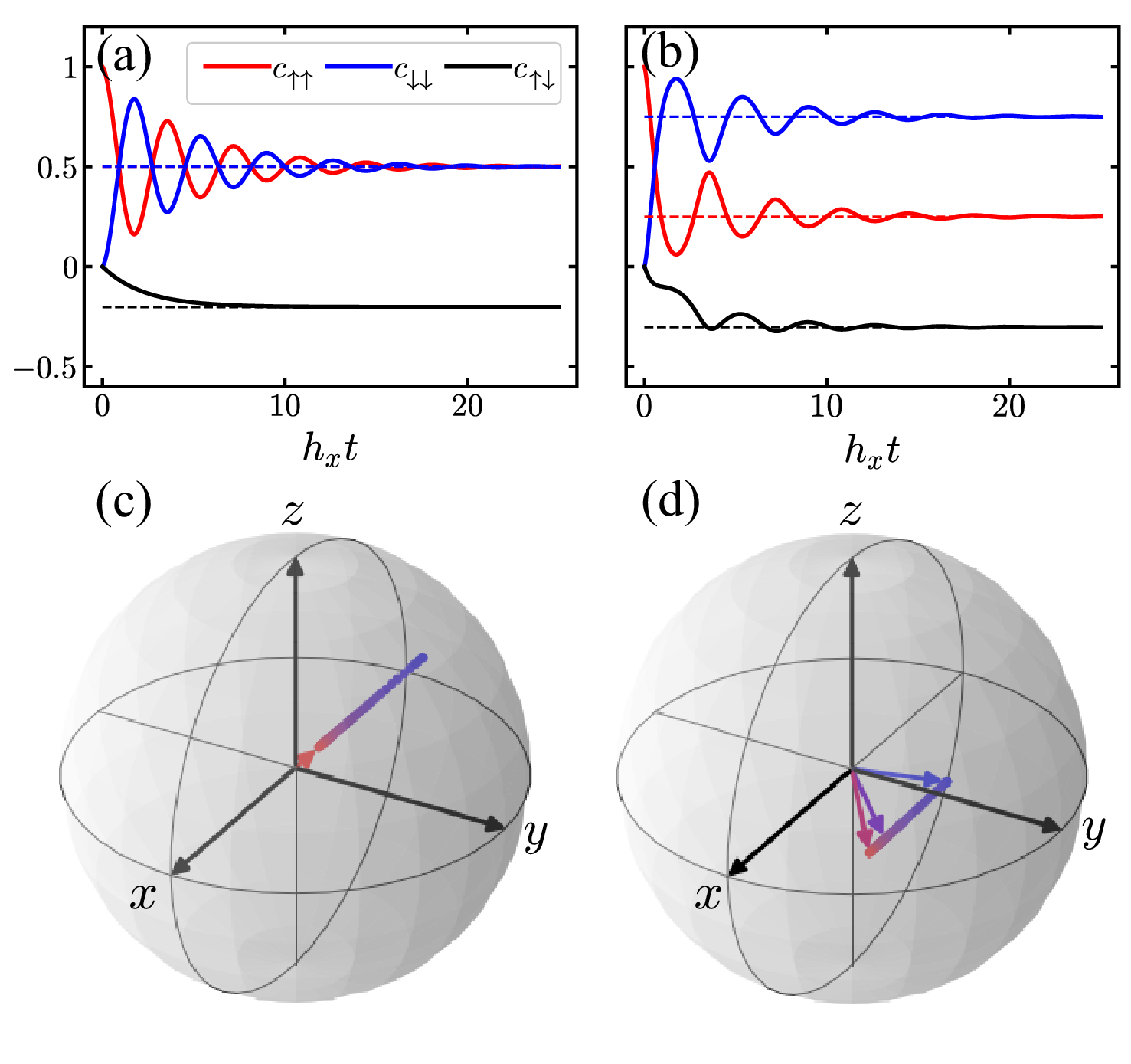}
\caption{\label{fig:fig2}
(Color online) The thermalization of the qubit~\eqref{eq:qubitmodel} is studied by analyzing the time evolution from an initial state $\ketbra{\uparrow}{\uparrow}$, using either (a) the biorthogonal GKSL master equation~\eqref{eq:bioGKSL} or (b) the right-state one~\eqref{eq:rightGKSL} at a temperature of $T/h_x=1$. We focus on the expansion coefficients $c_{\uparrow\uparrow}$ (red line), $c_{\downarrow\downarrow}$ (blue line) and $c_{\uparrow\downarrow}$ (black line) in the density matrix $\rho_\text{s}(t) = \sum_{\sigma,\sigma^\prime=\uparrow/\downarrow} c_{\sigma\sigma^\prime} (t) \ketbra{\sigma}{\sigma^\prime}$ as a function of the time $h_x t$. The dashed lines represent the anticipated coefficients in ETH. In the long term, under (c) BTE and (d) RTE, the vector $\braket{\bm \sigma}$ (points or vectors) is plotted in 3-dimensional coordinates as the temperature varies from $T=0.1$ (blue) to $5$ (red). Here, we use $h_y/h_x=0.5$ in the $\mathcal{PT}$-unbroken region.}
\end{center}
\end{figure}

\textit{A qubit.}
In the simplest example of the $\mathcal{PT}$-symmetric qubit~\cite{Naghiloo_2019}, the system Hamiltonian
\begin{equation}
\label{eq:qubitmodel}
H_\text{s} = h_x \sigma^x - i h_y \sigma^y
\end{equation}
has the biorthogonal eigenstates $\ket{\text{e}_\text{L}}$, $\ket{\text{e}_\text{R}}$, $\ket{\text{g}_\text{L}}$ and $\ket{\text{g}_\text{R}}$ for the excited state (e) and the ground state (g) with the energy $e_\text{e}$, $e_\text{g} = \pm\sqrt{h^2_x - h^2_y}$, respectively.
$\sigma^{x,y,z}$ denote the Pauli matrices in three axes and form a vector operator ${\bm \sigma} = (\sigma^x,\, \sigma^y,\, \sigma^z)^\intercal$.
We take $h_x=1$ as the energy unit and investigate the long-term behavior in the $\mathcal{PT}$-unbroken region $\vert h_y \vert<1$.
We choose $A = \sigma^z$, satisfying the thermalization condition~\eqref{eq:thermalizationCondition}.

After the initial preparation of a fully polarized pure state at temperature $T/h_x=1$, Figure~\ref{fig:fig2} shows the trajectory of the time-evolving matrix elements of $\rho_\text{s}$, according to Eq.~\eqref{eq:GKSL}.
When $h_x t \gg 1$, it can be easily verified that the stable steady state satisfies BBS~\eqref{eq:bioStatistics} in Fig.~\ref{fig:fig2}(a) and BRS~\eqref{eq:rightStatistics} in Fig.~\ref{fig:fig2}(b), neglecting the normalization coefficients.
\begin{figure}[t]
\begin{center}
\includegraphics[width=\columnwidth]{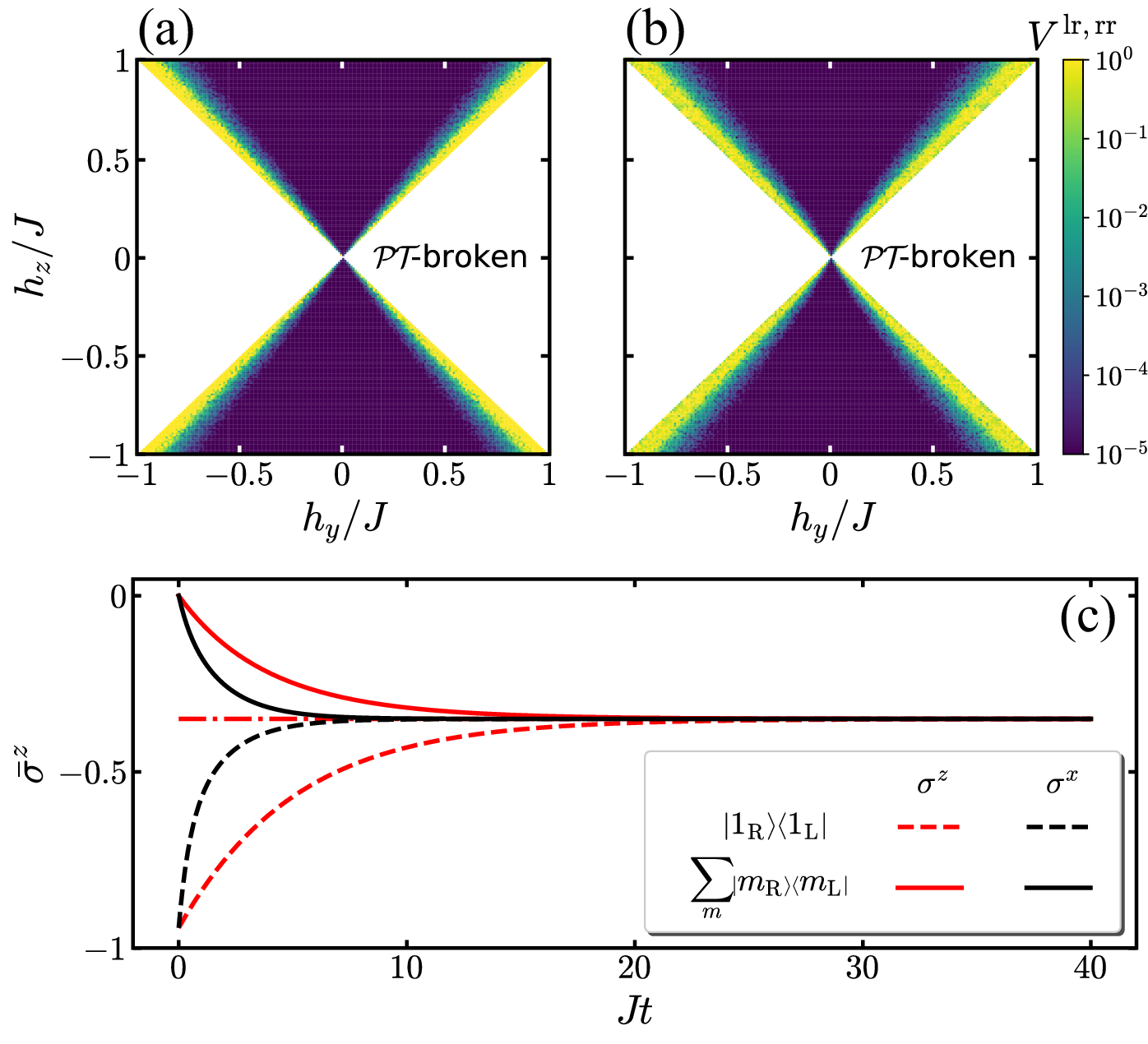}
\caption{\label{fig:fig3}
(Color online) In the $\mathcal{PT}$-unbroken region of the non-Hermitian Ising model~\eqref{eq:NHIsing}, we plot the variance of the long-term ($Jt \gg 1$) solution of either (a) the biorthogonal GKSL master equation~\eqref{eq:bioGKSL} and (b) the right-state one~\eqref{eq:rightGKSL} compared to the anticipated BBS~\eqref{eq:bioStatistics} and BRS \eqref{eq:rightStatistics}, respectively, in the ($h_y/J$, $h_z/J$)-plane. The variance near the exceptional point $\vert h_y \vert = \vert h_z \vert$ grows fast due to strong numerical instability. (c) When $h_y/J=0.2$ and $h_z/J=0.75$, we analyze the behavior of the averaged polarization in the $z$-axis $\bar{\sigma}^z = \sum^L_{\ell=1}\braket{\sigma^z_\ell} / L$ over time under BTE~\eqref{eq:rhoLR}, starting from either a mixed state at infinity temperature $\sum_m \ketbra{m_\text{R}}{m_\text{L}}$ (solid line) or a pure state at zero temperature$\ketbra{1_\text{R}}{1_\text{L}}$ (dashed line). We consider two cases of $A_\ell = \sigma^z_\ell$ (red) and $\sigma^x_\ell$ (black). The dashed line indicates the anticipated value according to ETH. Here, we set $L=4$ and $T/J=1$ for these calculations.
}
\end{center}
\end{figure}
In Figs.~\ref{fig:fig2}(c) and (d), we also plot the vector $\braket{\bm \sigma} = \text{tr}[\rho_\text{s} (t) {\bm \sigma}]_{h_x t \gg 1} = (\braket{\sigma^x},\, \braket{\sigma^y},\, \braket{\sigma^z})^\intercal$ in three dimensions. When we lower the temperature in Figs.~\ref{fig:fig2}(c), $\braket{\bm \sigma}$ under BTE~\eqref{eq:rhoLR} persists along $x$-axis in the equatorial plane, since $\text{Re} \braket{\sigma^{y, z}} = 0$. In contrast, under RTE~\eqref{eq:rhoRR}, the orientation of $\braket{\bm \sigma}$ continuously rotates in the equatorial plane.
It is worthy noticing that under BTE~\eqref{eq:rhoLR} the vector $\braket{\bm \sigma}$ can move out of the Bloch sphere at low temperatures because $\vert \braket{\text{e}_\text{L} \vert \sigma^x \vert \text{e}_\text{R}} \vert > 1$.

\textit{Non-Hermitian Ising model.}
In the model, we consider a real magnetic field $h_z$ in the $z$-axis and an imaginary one $i h_y$ in the $y$-axis. The latter can be implemented by the post-selection using Lindblad operators $A_\ell = \sqrt{h_z}(\sigma_\ell^x-i\sigma_\ell^y)$ for all sites~\cite{Deguchi2009, Yang2022}. Thus, the Hamiltonian of the system is defined as
\begin{eqnarray}
\label{eq:NHIsing}
H_\text{s} = \sum_{\ell=1}^{L-1} J \sigma_\ell^x \sigma_{\ell+1}^x + \sum_{\ell=1}^L (i h_y \sigma_\ell^y + h_z \sigma_\ell^z).
\end{eqnarray}
where the index $\ell$ ranges from $1$ to the size of the chain $L$, and the coupling strength $J=1$ sets the energy unit.
The operator $A_\ell$ is coupled to the individual heat bath at each site-$\ell$, i.e., $H_\text{sb} = \sum^L_{\ell=1} A_\ell \otimes B_\ell$, to avoid the (non-Markov) memory effects caused by the long-range correlations~\cite{Laine2012, Breuer2016}.
Although it is not a strict limitation, this assumption is physically plausible. More general forms of coupling could also be argued similarly, and the system will be thermalized if all the coupling operators $A_\ell$ satisfy the thermalization condition~\eqref{eq:thermalizationCondition}.

We study the thermalization of the system in the $\mathcal{PT}$-unbroken region $\vert h_z \vert > \vert h_y \vert$~\cite{Yang2022}.
To compare the solution of the GKSL master equation~\eqref{eq:GKSL} with BBS~\eqref{eq:bioStatistics} and BRS \eqref{eq:rightStatistics} at a typical temperature $T/J=1$, we investigate the variance of $V^{\text{lr},\text{rr}} = \Vert \rho_\text{s} (t) - \bar{\rho}^{\text{lr}, \text{rr}}_\text{s} \Vert_\infty$ as $Jt \gg 1$, where all density matrices have been normalized and $\Vert \cdots \Vert_\infty$ denotes the infinity-norm~\cite{Rajput2022}.
First, $A_\ell=\sigma_\ell^x$ has been taken in Fig~\ref{fig:fig3}(a) and (b) to ensure the system can be thermalized under BTE~\eqref{eq:rhoLR} and RTE~\eqref{eq:rhoRR} according to Eq.~\eqref{eq:thermalizationCondition}, even if more level spacing degeneracies ($M \gg 1$) exist and rich coherent transitions are allowed.
In the long term, the system can be described by density matrices~\eqref{eq:bioStatistics} and \eqref{eq:rightStatistics} within small error bars.
In Fig.~\ref{fig:fig3}(c), we also consider the case of $A_\ell=\sigma_\ell^z$, which violates the thermalization condition~\eqref{eq:thermalizationCondition}.
Under BTE~\eqref{eq:rhoLR}, this violation only leads to a longer period of decoherence.
In contrast, under RTE~\eqref{eq:rhoRR}, the thermalization cannot occur at all (not shown).

\begin{figure}[t]
\begin{center}
\includegraphics[width=\columnwidth]{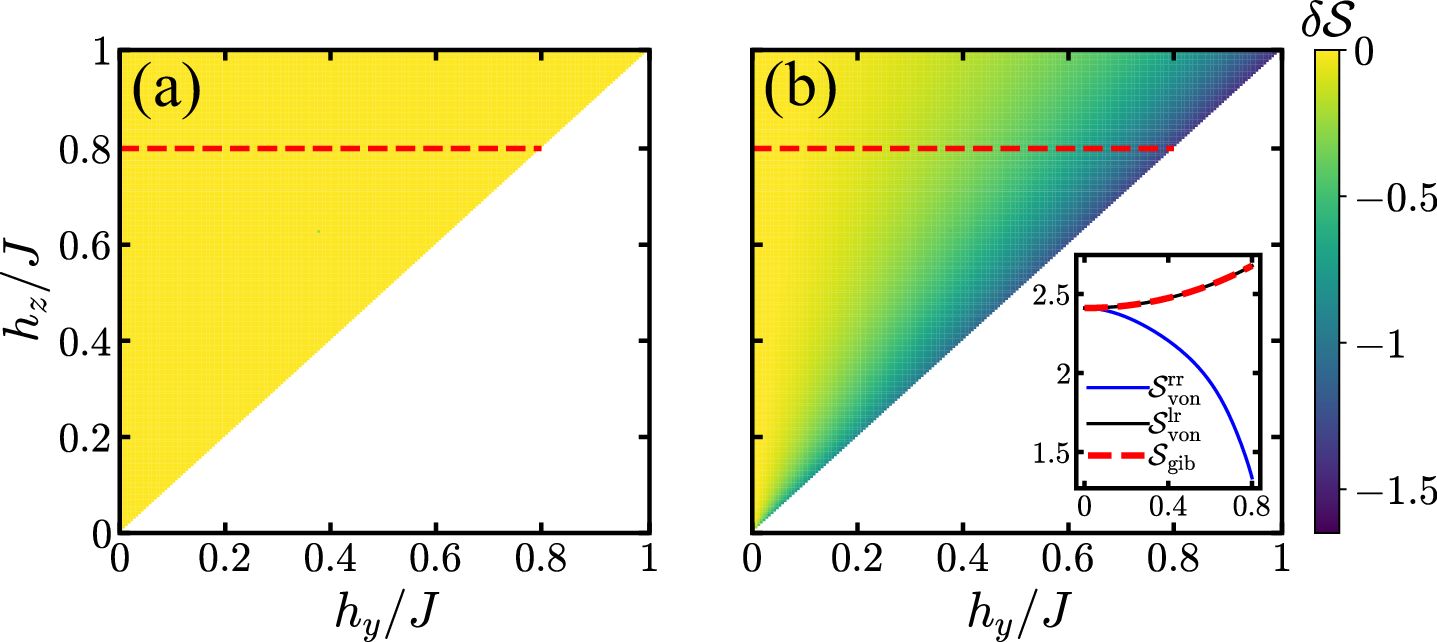}
\caption{\label{fig:fig4}
(Color online) After the thermalization of the non-Hermitian Ising model~\eqref{eq:NHIsing}, the discrepancy of the thermal entropies $\delta \mathcal{S} = \mathcal{S}_\text{von} - \mathcal{S}_\text{gib}$ is plotted in the ($h_y / J$, $h_z / J$)-plane. We consider (a) BBS~\eqref{eq:bioStatistics} and (b) BRS~\eqref{eq:rightStatistics}. Inset: Along a red dashed line $h_z/J=0.8$ and $h_y/J\in[0,\,0.8)$, we make a comparison of three thermal entropies: $\mathcal{S}^\text{lr}_\text{von}$ (black), $\mathcal{S}^\text{rr}_\text{von}$ (blue) and $\mathcal{S}_\text{gib}$ (red). Here, we set $L=4$ and $T/J=1$ for these calculations.}
\end{center}
\end{figure}

\textit{Thermal entropy.}
Two distinct statistics may bring us qualitative differences in measuring thermodynamic quantities. Again taking the non-Hermitian Ising model as an example, we analyzed the thermal von-Neumann entropy (vNE)~\cite{vonNeumann2018} of stable steady states described by density matrices~\eqref{eq:bioStatistics} and \eqref{eq:rightStatistics} after thermalization.
In Fig.~\ref{fig:fig4}(a), the biorthogonal vNE $\mathcal{S}^\text{lr}_\text{von}=-\text{tr} \bar{\rho}_\text{s}^\text{lr} \ln \bar{\rho}_\text{s}^\text{lr}$ is equal to the Gibbs entropy $\mathcal{S}_\text{gib} =-\sum^d_{m=1} \chi_m\log \chi_m$~\cite{Kardar2007}, trivially because of the biorthonormalization condition between the left and right eigenstates.
While in Fig.~\ref{fig:fig4}(b), we found that the right-eigenstate vNE $\mathcal{S}^\text{rr}_\text{von}=-\text{tr} \bar{\rho}_\text{s}^\text{rr}\ln 
\bar{\rho}_\text{s}^\text{rr}$ becomes lower than the Gibbs entropy $\mathcal{S}_\text{gib}$.
This phenomenon arises from the fact that the dissipation under the incomplete measurement lowers the information by reweighing bases or even erasing certain during the reorthogonalization of the right eigenstates in $\bar{\rho}_\text{s}^\text{rr}$, exemplified with the skin effects~\cite{Kawabata2023}.

\textit{Conclusion and discussions.}
We model the dynamics of non-Hermitian systems interacting with a bosonic heat bath in thermal equilibrium and establish the GKSL master equations using both the biorthogonal and the right-state time evolutions, where the latter has been obtained in previous studies~\cite{Du_2022, Cao_2023_1, Cao_2023_2}.
Our analysis reveals a sufficient condition~\eqref{eq:thermalizationCondition} for thermalization, ensuring that the long-term solutions of the equations exhibit BBS~\eqref{eq:bioStatistics} and BRS~\eqref{eq:rightStatistics}, respectively.
For a single qubit and a non-Hermitian Ising model in the $\mathcal{PT}$-unbroken region, we have shown that the density matrices agree with the anticipated statistics in the long term.

The early biorthogonal RMT suggested that the non-Hermitian system could generally not undergo decoherence due to the lack of entropic suppression in the off-diagonal terms of the biorthogonal measurement.
However, the biorthogonal GKSL master equation in this work indicates that a thermal state with BBS~\eqref{eq:bioStatistics} can be achieved once the system satisfies the thermalization condition~\eqref{eq:thermalizationCondition}.
This conflict may be attributed to a failure to consider the hidden relationship between the left and right eigenstates in the current version of RMT. Further investigation of the potential entropic effects on decoherence in the biorthogonal RMT is warranted.
Additionally, our work also extends the previously proposed BRS~\eqref{eq:rightStatistics} by adding a thermalization condition.

There are certain cases that may occur unexpectedly. For instance, if the thermalization condition~\eqref{eq:thermalizationCondition} is not met, we cannot guarantee that the diagonal density matrix survives in the long term. Additionally, the matrix $\mathbbm{A}$ may not be dense due to extra symmetries. In such cases, a thorough examination is necessary.

One more thing should be noted. If the right eigenstates do not satisfy the self-normalization condition, but obey the biorthonormalization condition with the left eigenvectors, the Pauli equation remains unchanged under BTE, and the system still evolves into the BBS~\eqref{eq:bioStatistics} when the thermalization condition~\eqref{eq:thermalizationCondition} is fulfilled.
In contrast, under RTE, different choices of right eigenstates yield different matrices $\mathbbm{L}$ in the Pauli master equation, which are connected by scaling transformations. Once the thermalization condition~\eqref{eq:thermalizationCondition} is satisfied for a choice, we have proved that BRS~\eqref{eq:rightStatistics} can be obtained in the long term following arbitrary $\mathbbm{L}$. An explanation is given in the Supplementary Material.

We are grateful to Su-Peng Kou for fruitful discussions.
This work is supported by grants: MOST~2022YFA1402700, NSFC~12174020, NSFC~12088101, NSFC~11974244 and NSAF~U2230402.
Computational resources from Tianhe-2JK at the Beijing Computational Science Research Center and Quantum Many-body \text{I} cluster at SPA, Shanghai Jiaotong University are also highly appreciated.

\bibliography{refs}
\end{document}

\newcommand{\ketbra}[2]{\mathinner{|{#1}\rangle \langle{#2}|}}

\title{\textbf{\large Supplementary Material for:\\ Diagnosing thermalization dynamics of non-Hermitian quantum systems via GKSL master equations}}

\author{Yiting Mao}
\affiliation{Beijing Computational Science Research Center, Beijing 100084, China}

\author{Peigeng Zhong}
\affiliation{Beijing Computational Science Research Center, Beijing 100084, China}

\author{Haiqing Lin}
\email[]{haiqing0@csrc.ac.cn}
\affiliation{Beijing Computational Science Research Center, Beijing 100084, China}
\affiliation{School of Physics, Zhejiang University, Hangzhou, 310058, China}

\author{Xiaoqun Wang}
\email[]{xiaoqunwang@zju.edu.cn}
\affiliation{School of Physics, Zhejiang University, Hangzhou, 310058, China}

\author{Shijie Hu}
\email[]{shijiehu@csrc.ac.cn}
\affiliation{Beijing Computational Science Research Center, Beijing 100084, China}
\affiliation{Department of Physics, Beijing Normal University, Beijing, 100875, China}

%\begin{abstract}
%
%\end{abstract}

\maketitle

\section{biorthogonal Pauli master equation}

In this section, we derive the biorthogonal Pauli master equation and discuss its two scenarios. We start our argument from Eqs.~(4) and (5) in the main text. We omit $\mathcal{L}_\text{s}[\rho_\text{s}]$ because it does not affect the steady state but only contributes a phase factor~\cite{Breuer_2007,BulnesCuetara2016}. Thus, $\mathcal{L}_\text{b}[\rho_\text{s}]$ can be rewritten as
\begin{equation}
\begin{split}
\mathcal{L}_\text{b}[\rho_\text{s}]=\frac{\text{d}}{\text{d}t}\rho_\text{s}&=\sum_{\omega\ne 0} \gamma(\omega)[(A_\omega \rho_\text{s} A_{-\omega})-\frac{1}{2}\{A_{-\omega}A_{\omega},\, \rho_\text{s}\}]\\
&=\underbrace{\sum_{m,n,m',n'}\delta_{\omega_{m'n'}=\omega_{mn}}\gamma(\omega_{m'n'})A_{\omega_{m'n'}}\rho_\text{s}A_{-\omega_{mn}}}_{\textcircled{1}}-\frac{1}{2}\underbrace{\sum_{m,n,m',n'}\delta_{\omega_{m'n'}=\omega_{mn}}\gamma(\omega_{m'n'})\{A_{-\omega_{m'n'}}A_{\omega_{mn}},\, \rho_\text{s}\}}_{\textcircled{2}}\, .
\end{split}
\end{equation}
We notice that the terms with $\omega_{mn}=0$ do not contribute to the effective physical processes and can be neglected in the following argument. After we define $c_{mn}(t)=\bra{m_\text{L}}\rho_\text{s}(t)\ket{n_\text{R}}$, $\textcircled{1}$ and $\textcircled{2}$ can be expressed as
\begin{equation}
\begin{split}
\textcircled{1}=\sum_{m,n,m',n'}\delta_{\omega_{m'n'}=\omega_{mn}}\gamma(\omega_{m'n'})&\ket{m'_\text{R}}\bra{m'_\text{L}}A\ket{n'_\text{R}}\bra{n'_\text{L}}\rho_\text{s}\ket{n_\text{R}}\bra{n_\text{L}}A\ket{m_\text{R}}\bra{m_\text{L}}\, ,
\\
\textcircled{2}=\sum_{m,n,m',n'}\delta_{\omega_{m'n'}=\omega_{mn}}\gamma(\omega_{m'n'}) &\bigg[ \ket{n'_\text{R}}\bra{n'_\text{L}}A\ket{m'_\text{R}}\braket{m^\prime_\text{L} \vert m_\text{R}}\bra{m_\text{L}}A\ket{n_\text{R}}\bra{n_\text{L}}\rho_\text{s}\\
&+\rho_\text{s} \ket{n'_\text{R}}\bra{n'_\text{L}}A\ket{m'_\text{R}}\braket{m'_\text{L} \vert m_\text{R}}\bra{m_\text{L}}A\ket{n_\text{R}}\bra{n_\text{L}} \bigg]\, .
\end{split}
\end{equation}
Considering $\rho_\text{s}^{(\Delta)}(t)=\sum_{p=1}^Mc_p(t)\ket{(m_p)_\text{R}}\bra{(n_p)_\text{L}}$ with bias $\Delta=e_{n_p}-e_{m_p}$ and $c_p(t)=c_{m_pn_p}(t)$, $\mathcal{L}_\text{b}[\rho_\text{s}^{(\Delta)}]$ is given by
\begin{equation}
\label{elements}
\frac{\text{d}}{\text{d}t}c_p=\frac{\text{d}}{\text{d}t}\bra{(m_p)_\text{L}}\rho_\text{s}^{(\Delta)}\ket{(n_p)_\text{R}}=\bra{(m_p)_\text{L}}\textcircled{1}\ket{(n_p)_\text{R}}-\frac{1}{2}\bra{(m_p)_\text{L}}\textcircled{2}\ket{(n_p)_\text{R}}\, .
\end{equation}
We calculate them individually, that is,
\begin{eqnarray}
\begin{split}
\bra{(m_p)_\text{L}}\textcircled{1}\ket{(n_p)_\text{R}}&=\sum_{m,n,m',n'}\delta_{\omega_{m'n'}=\omega_{mn}}\gamma(\omega_{m'n'})\mathbbm{A}_{m'n'}\mathbbm{A}_{nm}c_{n'n}\delta_{m_pm'}\delta_{mn_p}\\
&=\sum_{n,n'}\delta_{\omega_{m_pn'}=\omega_{n_pn}}\gamma(\omega_{m_pn'})\mathbbm{A}_{m_pn'}\mathbbm{A}_{nn_p}c_{n'n}=\sum_{m',n'}\delta_{\omega_{m_pm'}=\omega_{n_pn'}}\gamma(\omega_{m_pm'})\mathbbm{A}_{m_pm'}\mathbbm{A}_{n'n_p}c_{m'n'}\, ,\\
\bra{(m_p)_\text{L}}\textcircled{2}\ket{(n_p)_\text{R}}&=\sum_{m,n,m',n'}\delta_{\omega_{m'n'}=\omega_{mn}}\gamma(\omega_{m'n'})\mathbbm{A}_{n'm'}\mathbbm{A}_{mn}c_{nn_p}+\delta_{\omega_{m'n'}=\omega_{mn}}\gamma(\omega_{m'n'})\mathbbm{A}_{n'm'}\mathbbm{A}_{mn}c_{m_pn'}\delta_{mm'}\delta_{nn_p}\\
&=\sum_{m',n}\delta_{\omega_{m'm_p}=\omega_{m'n}}\gamma(\omega_{m'm_p})\mathbbm{A}_{m_pm'}\mathbbm{A}_{m'n}c_{nn_p}+\sum_{m',n'}\delta_{\omega_{m'n'}=\omega_{m'n_p}}\gamma(\omega_{m'n'})\mathbbm{A}_{n'm'}\mathbbm{A}_{m'n_p}c_{m_pn'}\\
&=\sum_{m',n'} \bigg[\delta_{\omega_{m'm_p}=\omega_{m'n'}}\gamma(\omega_{m'm_p})\mathbbm{A}_{m_pm'}\mathbbm{A}_{m'n'}c_{n'n_p}+\delta_{\omega_{m'n'}=\omega_{m'n_p}}\gamma(\omega_{m'n'})\mathbbm{A}_{n'm'}\mathbbm{A}_{m'n_p}c_{m_pn'} \bigg]\, ,
\end{split}\ 
\end{eqnarray}
where $\mathbbm{A}_{mn}=\bra{m_\text{L}}A\ket{n_\text{R}}$.

We analyze the cases of $\Delta=0$ and $\Delta\ne 0$ in Eq.~\eqref{elements} separately. 
\begin{itemize}
\item [(\textbf{i})] If $\Delta=0$ $(m_p=n_p)$~\cite{BulnesCuetara2016}, then $M$ is equal to the dimension $d$ of $H_\text{s}$. Consequently, one can obtain
\begin{equation}
\label{diagonal1}
\begin{split}
\frac{\text{d}}{\text{d}t}c_{p}=&\sum_{m',n'}\delta_{\omega_{m_pm'}=\omega_{m_pn'}}\gamma(\omega_{m_pm'})\mathbbm{A}_{m_pm'}\mathbbm{A}_{n'm'}c_{m'n'}\\
&-\frac{1}{2}\sum_{m',n'}\bigg[\delta_{\omega_{m'm_p}=\omega_{m'n'}}\gamma(\omega_{m'm_p})\mathbbm{A}_{m_pm'}\mathbbm{A}_{m'n'}c_{n'm'}+\delta_{\omega_{m'n'}=\omega_{m'm_p}}\gamma(\omega_{m'n'})\mathbbm{A}_{n'm'}\mathbbm{A}_{m'm_p}c_{m_pn'}\bigg]\\
=&\sum_{n'}\gamma(\omega_{m_pn'})\mathbbm{A}_{m_pn'}\mathbbm{A}_{n'm_p}c_{n'} -\frac{1}{2}\sum_{m'}\bigg[\gamma(\omega_{m'm_p})\mathbbm{A}_{m_pm'}\mathbbm{A}_{m'm_p}c_{p}+\gamma(\omega_{m'm_p})\mathbbm{A}_{m_pm'}\mathbbm{A}_{m'm_p}c_{p}\bigg]\, .
\end{split}
\end{equation}
We replace the indices $m',n'$ with $q$, $m_p$ with $p$ here, then Eq.~\eqref{diagonal1} can be rewritten as
\begin{equation}
\label{diagonal2}
\frac{\text{d}}{\text{d}t}c_{p}=\sum_{q\ne p}[\gamma(\omega_{pq})\kappa_{qp}c_{q}-\gamma(\omega_{qp})\kappa_{pq}c_{p}]
\end{equation}
with the transition rate $\kappa_{pq}=\mathbbm{A}_{pq}\mathbbm{A}_{qp}=\kappa_{qp}$. By defining ${\bm c}=(c_1,\, \cdots,\, c_M)^\intercal$, Eq.~\eqref{diagonal2} can be written as the Pauli master equation~\cite{Pauli28},
\begin{equation}
\label{Paulieq}
\frac{\text{d}}{\text{d}t}{\bm c}=\mathbb{L}{\bm c}\, ,
\end{equation}
where the diagonal elements and off-diagonal elements of $\mathbb{L}$ are
\begin{equation}
\mathbb{L}_{pp} = -\sum_{q\ne p}\gamma(\omega_{qp})\kappa_{pq}\, ,\qquad \mathbb{L}_{qp}=\gamma(\omega_{qp})\kappa_{pq}\, .
\end{equation}
It is easy to verify that the non-Hermitian matrix $\mathbb{L}$ has at least one left eigenvector $\bar{\bm c}_\text{L}=(1,\, \cdots,\, 1)^\intercal$, corresponding to the eigenvalue $\lambda=0$,
\begin{equation}
\begin{split}
\bar{\bm c}_\text{L}^\intercal \mathbb{L}&=(1,\, \cdots,\, 1)\begin{pmatrix}
&-\sum_{q\ne 1}\gamma(\omega_{q1})\kappa_{1q} & \gamma(\omega_{12})\kappa_{21} & \cdots &\gamma(\omega_{1M})\kappa_{M1}\\
&\gamma(\omega_{21})\kappa_{12} & -\sum_{q\ne 2}\gamma(\omega_{q2})\kappa_{2q} & \cdots &\gamma(\omega_{2M})\kappa_{M2}\\
&\vdots &\vdots &\ddots &\vdots\\
&\gamma(\omega_{M1})\kappa_{1M} & \gamma(\omega_{M2})\kappa_{2M} & \cdots & -\sum_{q\ne M}\gamma(\omega_{qM})\kappa_{Mq}
\end{pmatrix}\\
&=\left(\sum_{q=1}^M\mathbb{L}_{q1},\, \cdots,\, \sum_{q=1}^M\mathbb{L}_{qM}\right)=(0,\, \cdots,\, 0)=\lambda\bar{\bm c}^\intercal_\text{L}\, .
\end{split}
\end{equation}
Therefore, the corresponding right eigenvector $\bar{\bm c}$, which shares $\lambda$ with $\bar{\bm c}_\text{L}$, survives as the stable steady state in the long term.

\item [(\textbf{ii})] It is more difficult to argue the case of a certain $\Delta\ne 0$ $(m_p\ne n_p)$ in Eq.~\eqref{elements}. Similar to Eq.~\eqref{diagonal1}, we can get
\begin{equation}
\label{off-diagonal1}
\begin{split}
\frac{\text{d}}{\text{d}t}c_p=\sum_{m',n'}& \bigg[\delta_{\omega_{m_pm'}=\omega_{n_pn'}}\gamma(\omega_{m_pm'})\mathbbm{A}_{m_pm'}\mathbbm{A}_{n'n_p}c_{m'n'}\\
&-\frac{1}{2}\delta_{\omega_{m'm_p}=\omega_{m'n'}}\gamma(\omega_{m'm_p})\mathbbm{A}_{m_pm'}\mathbbm{A}_{m'n'}c_{m'n_p}-\frac{1}{2}\delta_{\omega_{n'm'}=\omega_{n'n_p}}\gamma(\omega_{n'n_p})\mathbbm{A}_{m'n'}\mathbbm{A}_{n'n_p}c_{m_pm'}\bigg]\, .
\end{split}
\end{equation}
To obtain nonzero terms in Eq.~\eqref{off-diagonal1}, we take $n'=m_p$ and $m'=n_p$ in the second and the third terms, respectively, and consider $\omega_{m_pm_q}=\omega_{n_pn_q}$ in the first term when $M\ge 2$. We thus have
\begin{equation}
\label{off-diagonal evolution1}
\begin{split}
\frac{\text{d}}{\text{d}t}c_p=&\sum_{q\ne p}\gamma(\omega_{m_pm_q})\mathbbm{A}_{m_pm_q}\mathbbm{A}_{n_qn_p}c_{q}-\frac{1}{2}\sum_{q\ne p}\gamma(\omega_{m_qm_p})\kappa_{m_qm_p}c_{p}-\frac{1}{2}\sum_{q\ne p}\gamma(\omega_{n_qn_p})\kappa_{n_qn_p}c_{p}\\
&-\frac{1}{2}\sum_{m'\notin{\bm v}_m}\gamma(\omega_{m'm_p})\kappa_{m_pm'}c_{p}-\frac{1}{2}\sum_{n'\notin{\bm v}_n}\gamma(\omega_{n'n_p})\kappa_{n'n_p}c_{p}
\end{split}
\end{equation}
with ${\bm v}_m=\{m_1,\,\cdots,\,m_M\}$, ${\bm v}_{n}=\{n_1,\,\cdots,\,n_M\}$ representing the sets of the available pairs of $m$ and $n$. ${\bm c}$ also obeys the Pauli master equation Eq.~\eqref{Paulieq}. The matrix elements of $\mathbb{L}$ are given by
\begin{equation}
\begin{split}
\mathbb{L}_{pp}=&-\frac{1}{2}\left[\sum_{m'\ne m_p}\gamma(\omega_{m'm_p})\kappa_{m_pm'}+\sum_{n'\ne n_p}\gamma(\omega_{n'n_p})\kappa_{n_pn'}\right]=\mathbb{L}_{pp}^{(\text{i})}+\mathbb{L}_{pp}^{(\text{o})}\, ,\qquad \mathbb{L}_{qp}=\gamma(\omega_{m_qm_p})\mathbbm{A}_{m_qm_p}\mathbbm{A}_{n_pn_q}\, ,\\
\mathbb{L}_{pp}^{(\text{i})}=&-\sum_{q\ne p}\gamma(\omega_{m_qm_p})(\kappa_{m_pm_q}+\kappa_{n_pn_q})/2\, ,\qquad \mathbb{L}_{pp}^{(\text{o})}=-\frac{1}{2}\sum_{m'\notin{\bm v}_m}\gamma(\omega_{m'm_p})\kappa_{m_pm'}-\frac{1}{2}\sum_{n'\notin{\bm v}_n}\gamma(\omega_{n'n_p})\kappa_{n_pn'}\, .
\end{split}
\end{equation}
Here we divide $\mathbb{L}_{pp}$ into the part $\mathbb{L}_{pp}^{(\text{i})}$ in the sector and $\mathbb{L}_{pp}^{(\text{o})}$ out of the sector.
If $\mathbb{L}$ is a strictly diagonally dominant matrix with negative diagonal elements, i.e., $|\mathbb{L}_{pp}|-\sum_{q\ne p}|\mathbb{L}_{qp}|>0$ and $|\mathbb{L}_{pp}|<0$ for arbitrary $p$, the real parts of eigenvalues of $\mathbb{L}$ are all negative~\cite{ger1931,Timm2009}, which implies that the off-diagonal density matrix $\rho_\text{s}^{(\Delta\ne 0)}$ vanishes in the long term.

To prove that $\mathbb{L}$ is strictly diagonally dominant, we calculate
\begin{equation}
\label{inequality}
\begin{split}
&|\mathbb{L}_{pp}|-\sum_{q\ne p}|\mathbb{L}_{qp}|=|\mathbb{L}_{pp}^{(\text{i})}|+|\mathbb{L}_{pp}^{(\text{o})}|-\sum_{q\ne p}|\mathbb{L}_{qp}|
\\
=&\frac{1}{2}\left|\sum_{q\ne p}\gamma(\omega_{m_qm_p})(\kappa_{m_qm_p}+\kappa_{n_qn_p})+\sum_{m'\notin{\bm v}_m}\gamma(\omega_{m'm_p})\kappa_{m'm_p}+\sum_{n'\notin{\bm v}_n}\gamma(\omega_{n'n_p})\kappa_{n_pn'}\right|\\
&-\sum_{q\ne p}\gamma(\omega_{m_qm_p})|\mathbbm{A}_{m_qm_p}\mathbbm{A}_{n_pn_q}|
\end{split}
\end{equation}
where $|\cdots|$ denotes the norm. Under the thermalization condition Eq.~(7) in the main text, it can be proved that $\kappa_{m_qm_p} \ge 0$ and $\kappa_{n_qn_p} \ge 0$, so that Eq.~\eqref{inequality} can be rewritten as
\begin{eqnarray}
\label{diagonally dominant}
|\mathbb{L}_{pp}|-\sum_{q\ne p}|\mathbb{L}_{qp}| &= &\frac{1}{2}\left[\sum_{q\ne p}\gamma(\omega_{m_qm_p})(|\mathbbm{A}_{m_pm_q}|-|\mathbbm{A}_{n_pn_q}|)^2+\sum_{m'\notin{\bm v}_m}\gamma(\omega_{m'm_p})|\kappa_{m'm_p}|+\sum_{n'\notin{\bm v}_n}\gamma(\omega_{n'n_p})|\kappa_{n_pn'}|\right] \nonumber\\
&>& |\mathbb{L}_{pp}^{(\text{i})}|-\sum_{q\ne p}|\mathbb{L}_{qp}|=\frac{1}{2}\left[\sum_{q\ne p}\gamma(\omega_{m_qm_p})(|\mathbbm{A}_{m_pm_q}|-|\mathbbm{A}_{n_pn_q}|)^2\right]\ge 0\, .
\end{eqnarray}
Moreover, the thermalization condition also ensures $\kappa_{m'm_p}\ge0$ and $\kappa_{n_pn'}\ge0$, implying that the diagonal elements $\mathbb{L}_{pp}\le0$ for arbitrary $p$. All inequalities become strict as $\mathbbm{A}$ is dense without any extra symmetries.

On the one hand, if $|\mathbb{L}_{pp}^{(\text{i})}|-\sum_{q\ne p}|\mathbb{L}_{qp}|>0$ for the specified $p$, the coherent transitions from the same initial projector $\ket{(m_p)_\text{R}}\bra{(n_p)_\text{L}}$ to various final projectors $\ket{(m_q)_\text{R}}\bra{(n_q)_\text{L}}$ yield the reduction of $\vert{\bm c}\vert$. On the other hand, only $c_p$ decays due to the dissipation term originating from the energy levels out of the sector when $\mathbb{L}_{pp}^{(\text{o})} < 0$. Either of the above mechanisms can give rise to the vanishing of $\rho_\text{s}^{(\Delta\ne 0)}$ in the long term for $M\ge2$.

In particular, if we cannot find a $q$ to satisfy $\omega_{m_qm_p}=\omega_{n_qn_p}$, i.e. $M=1$, quantum jumps are absent and Eq.~\eqref{off-diagonal1} becomes 
\begin{equation}
\label{off-diagonal evolution2}
\frac{\text{d}}{\text{d}t}c_p=-\frac{1}{2}\sum_{m'}\gamma(\omega_{m'm_p})\kappa_{m_pm'}c_p-\frac{1}{2}\sum_{n'}\gamma(\omega_{n_pn'})\kappa_{n'n_p}c_p\, .
\end{equation}
The dissipation term in Eq.~\eqref{off-diagonal1} still causes the vanishing of $\rho_\text{s}^{(\Delta\ne 0)}$, as both $\kappa_{m'm_p}$ and $\kappa_{n_pn'}$ are non-negative under the thermalization condition.
\end{itemize}

Overall, the thermalization condition guarantees that only $\rho_\text{s}^{(0)}(t)$ can survive in the long term.

\section{Boltzmann right-eigenstate statistics}
In this section, we derive the Boltzmann right-eigenstate statistics (BRS) under the right-state time evolution (RTE). We assume that the system Hamiltonian $H_\text{s}$ has the self-normalized right eigenstates $\{\ket{m_\text{R}}\}$ with the real energy $e_1>e_2>\cdots>e_d$. Given the density matrix in diagonal $\rho^{(0)}_\text{s}(t)=\sum^d_{m=1} c_m(t)\ketbra{m_\text{R}}{m_\text{R}}$, we can rewrite Eq.~(10) in the main text as
\begin{equation}
\mathcal{L}_\text{s}[\rho^{(0)}_\text{s}]=0\, ,\qquad \mathcal{L}_\text{d}[\rho^{(0)}_\text{s}]=\sum_{m\ne n}\gamma(\omega_{mn})\{A_{nm}A_{mn},\, \rho^{(0)}_\text{s}\}_\dagger\, ,\qquad\mathcal{L}_\text{j}[\rho^{(0)}_\text{s}]=\sum_{m\ne n}2\gamma(\omega_{mn})(A_{mn}\rho^{(0)}_\text{s}(A_{mn})^\dagger)\,
\end{equation}
where $\omega_{mn}=e_n-e_m$ and the spectrum function $\gamma$ keeps the KMS condition $\gamma(\omega_{nm})/\gamma(\omega_{mn})=e^{-\beta\omega_{mn}}$ at temperature $T=1/\beta$.
Then, the GKSL master equation Eq.~(4) in the main text under RTE becomes
\begin{equation}
\mathcal{L}[\rho^{(0)}_\text{s}] = \mathcal{L}_\text{b}[\rho^{(0)}_\text{s}] = \sum_{m>n}(\gamma(\omega_{nm})\tilde{W}_{mn}+\gamma(\omega_{mn})\tilde{W}_{nm})\, ,
\end{equation}
where
\begin{equation}
\begin{split}
&\tilde{W}_{mn} = \left(\mathbbm{A}_{nm}\mathbbm{A}_{nm}^*\ketbra{n_\text{R}}{n_\text{R}}-\mathbbm{A}_{mn}^*\mathbbm{A}_{nm}^*\ketbra{m_\text{R}}{m_\text{R}}\right)c_{m}(t)\, ,\\
&\tilde{W}_{nm} = \left(\mathbbm{A}_{mn}\mathbbm{A}_{mn}^*\ketbra{m_\text{R}}{m_\text{R}}-\mathbbm{A}_{nm}\mathbbm{A}_{mn}\ketbra{n_\text{R}}{n_\text{R}}\right)c_{n}(t)\, .
\end{split}
\end{equation}

We can consider a simplified scenario where only the $m$-th and $n$-th states ($e_m > e_n$) are involved, i.e., $\rho^{(0)}_\text{s}=c_m\ketbra{m_\text{R}}{m_\text{R}}+c_n\ketbra{n_\text{R}}{n_\text{R}}$, then we have
\begin{equation}
\label{Right_Pauli_master_equation}
\frac{\text{d}}{\text{d}t}
\begin{pmatrix}
c_m\\
\\
c_n
\end{pmatrix}=
\begin{pmatrix}
-\gamma(\omega_{nm})\mathbbm{A}_{mn}^*\mathbbm{A}_{nm}^* & \gamma(\omega_{mn})\mathbbm{A}_{mn}\mathbbm{A}_{mn}^*\\
\\
\gamma(\omega_{nm})\mathbbm{A}_{nm}\mathbbm{A}_{nm}^* & -\gamma(\omega_{mn})\mathbbm{A}_{nm}\mathbbm{A}_{mn}
\end{pmatrix}\begin{pmatrix}
c_m\\
\\
c_n
\end{pmatrix}=\mathbb{L}\begin{pmatrix}
c_m\\
\\
c_n
\end{pmatrix}\, .
\end{equation}
When the thermalization condition Eq.~(7) in the main text is satisfied, then Eq.~\eqref{Right_Pauli_master_equation} becomes
\begin{equation}
\label{Right_Pauli_master_equation_obeyTC}
\frac{\text{d}}{\text{d}t}
\begin{pmatrix}
c_m\\
\\
c_n
\end{pmatrix}=
\kappa_{mn}\begin{pmatrix}
-\gamma(\omega_{nm}) & \gamma(\omega_{mn})\\
\\
\gamma(\omega_{nm}) & -\gamma(\omega_{mn})
\end{pmatrix}\begin{pmatrix}
c_m\\
\\
c_n
\end{pmatrix}\, .
\end{equation}
For the matrix $\mathbb{L}$, an eigenstate $\bar{\bm c}=(\alpha^2e^{\beta\omega_{mn}},\, 1)^\intercal$ corresponds to the maximum eigenvalue $\lambda=0$. In the long term, only this eigenstate survives, giving the Boltzmann distribution for two levels, that is,
\begin{equation}
\label{TLBoltzmann_distribution}
\bar{\rho}_\text{s}=\frac{e^{\beta\omega_{mn}}\ketbra{m_\text{R}}{m_\text{R}}+\ketbra{n_\text{R}}{n_\text{R}}}{1+e^{\beta\omega_{mn}}}\, .
\end{equation}
 
We can change the norm of the $n$-th right eigenstate, i.e., $\ket{n_\text{R}'}=\alpha\ket{n_\text{R}}$, but scale the corresponding left eigenstate to the revised one $\ket{n_\text{L}'}=(1/ \alpha)\ket{n_\text{L}}$ to preserve the validity of the normalization condition of $\braket{n_\text{L}' \vert n_\text{R}'} = 1$, where $\alpha$ is a positive real number. Then, Eq.~\eqref{Right_Pauli_master_equation} can be rewritten as
\begin{equation}
\frac{\text{d}}{\text{d}t}
\begin{pmatrix}
c_m\\
\\
c_n
\end{pmatrix}=
\kappa_{mn}\begin{pmatrix}
-\gamma(\omega_{nm}) & \alpha^2\gamma(\omega_{mn})\\
\\
\frac{1}{\alpha^2}\gamma(\omega_{nm}) & -\gamma(\omega_{mn})
\end{pmatrix}\begin{pmatrix}
c_m\\
\\
c_n
\end{pmatrix}\, ,
\end{equation}
in the new eigenstate set.
We can see that $\mathbb{L}$ still has an eigenstate $\bar{\bm c}=(\alpha^2e^{\beta\omega_{mn}},\, 1)^\intercal$ with the maximum eigenvalue $\lambda=0$, which also yields the Boltzmann distribution $\bar{\rho}'_\text{s}$, the same as Eq.~\eqref{TLBoltzmann_distribution} after it is normalized. In detail, we get
\begin{equation}
\bar{\rho}'_\text{s}=\frac{\alpha^2e^{\beta\omega_{mn}}\ketbra{m_\text{R}'}{m_\text{R}'}+\ketbra{n_\text{R}'}{n_\text{R}'}}{\alpha^2e^{\beta\omega_{mn}}\braket{m_\text{R}'|m_\text{R}'}+\braket{n_\text{R}'|n_\text{R}'}}=\bar{\rho}_\text{s}
\end{equation}
Similarly, this conclusion can be extended to the case of multiple energy levels. As a result, the system can evolve into a thermal state described by BRS under BTE.

\bibliography{sm_refs}